\documentclass[%
 reprint,
 amsmath,amssymb,
 aps,
]{revtex4-1}
\usepackage{graphicx}% Include figure files
\usepackage{dcolumn}% Align table columns on decimal point
\usepackage{bm}% bold math
\begin{document}
%\graphicspath{{Images/}}
\preprint{APS/123-QED}

\title{Analytical expression for radial distribution function of hard sphere system: \\density derivative and application to perturbation theories}% Force line breaks with \\

\author{Timur Aslyamov}
 \email{t.aslyamov@gmail.com}
 \affiliation{Schlumberger Moscow Research Center; 13, Pudovkina str., Moscow 119285, Russia}%Lines break automatically or can be forced with \\

\date{\today}% It is always \today, today,

\begin{abstract}
Model of hard sphere system is important part of modern theories of liquids. Radial distribution function of hard sphere fluid represented in form of explicit analytical expression allows to obtain thermodynamic potentials in analytical form, which is very helpful for further analysis and applications. A new analytical expression of radial distribution function of hard sphere fluid is developed. The density derivative of the radial distribution function and the first two terms of Barker-Henderson perturbation theory are derived as analytical expressions. The obtained results agree well with published simulation data. These results have important implications for real fluid modeling using density functional theory and perturbation theories.
%\begin{description}
%\end{description}
\end{abstract}

\pacs{Valid PACS appear here}% PACS, the Physics and Astronomy
%Use showkeys class option if keyword
\maketitle

%\tableofcontents
\section{Introduction}
Perturbation theories (PT) play crucial role in description of thermodynamic properties of fluids at wide range of densities \cite{zwanzig1954high, weeks1971role, barker1967perturbation1, barker1967perturbation2}. General idea of these theories is the transition from complex real system to more simple reference system, which is described by the repulsive part of interaction potential. As the reference system the hard sphere (HS) fluid is commonly used  \cite{hansen1990theory}. 

One of the most successful PT is the one suggested by Barker-Henderson (BH) \cite{barker1967perturbation1, barker1967perturbation2}. This theory describes molecules as spheres of diameter $\sigma$ interacting by pair potential $U(r)$, where $r$ is the distance between centers of the molecules. In BH theory the intermolecular potential is decomposed into a sum of two piecewise functions  $U(r)=U_0(r)+U_1(r)$: a reference $U_0$ and a perturbation $U_1$ term. Then, it is possible to represent the Helmholtz free energy $A$ as following series
$$ 
A=\sum_{n=0}^{\infty}\beta^n A_n,
$$
where $\beta=(kT)^{-1}$, $k$ is the Boltzmann constant, $T$ is temperature, $A_0$ is the Helmholtz free energy of the reference fluid, $A_{n}$ is the n-th order perturbation term. Exact properties of  $A_0$ are unknown, however, it is  possible to map the reference fluid to hard spheres system with another effective diameter \cite{barker1967perturbation2}:
$$
d=\int_{0}^{\sigma}\left(1-e^{-\beta U(r)}\right)dr<\sigma.
$$
In order to describe the behavior of reference system, the theory of correlation functions is used \cite{hansen1990theory, barker1976liquid}. Thus, the only data one needs for calculation of the first term of PT $a_1=\beta A_1/N$ ($N$ is the number of molecules) is information about radial distribution function $g(r)$ (RDF) (i.e., pair correlation function):

\begin{eqnarray}
\label{a1}
a_1=2\pi\rho \int_{\sigma}^{\infty}g(r)U(r)r^2dr.
\end{eqnarray}
Description of second-order perturbation term is more complicated, since information about correlation functions of order higher then the first is needed \cite{smith1970approximate}. However, Barker and Henderson developed compressibility approximation \cite{barker1967perturbation1, barker1967perturbation2}. According to this model the second term $a_2=\beta A_2/N$ is:
\begin{eqnarray}
\label{a2}
a_2=-\pi\rho K^{HS}\int_{\sigma}^{\infty}g(r)(U(r))^2r^2dr,
\end{eqnarray}
where $K^{HS}=kT(\partial \rho/ \partial P)_T$ is the isothermal compressibility of reference system (HS fluid), $\rho$ and $P$ are the density and the pressure of HS fluid, respectively. 
Parameter $K^{HS}$ can be calculated from the density derivative of the Carnahan Starling compressibility \cite{carnahan1969equation}
$$
K^{HS}=\frac{(1-\phi)^4}{1+4\phi+4\phi^2-4\phi^3+\phi^4},
$$
where $\phi=\pi\rho d^3/6$ is the dimensionless density. Everywhere below when we say the density $\phi$.  As one can see above, all the required properties of HS system can be obtained from RDF.

The initial description of HS fluid was performed by by Wertheim \cite{wertheim1963exact} and Thile \cite{thiele1963equation}. They obtained the solution of Ornstein–Zernike  equation in Percus-Yevick (PY) approximation \cite{percus1958analysis}. However, analytical result was obtained only for the Laplace transform $G(t, \phi)$ of product $r g(r, \phi)$ 
\begin{eqnarray}
G(t,\phi)=\int_{0}^{\infty}rg(r,\phi)e^{-t r}dr.
\end{eqnarray}
Wertheim \cite{wertheim1963exact} obtained Laplace image $G(t)$ as analytical function. According to him RDF has the following form:
\begin{align}
	\label{RDF_Wertheim}
	g(r,\phi)=\frac{1}{2\pi i}	\int_{\delta-i\infty}^{\delta+ i \infty}\frac{t L(t,\phi)e^{t r}dt}{12\phi r \left[L(t, \phi)+S(t, \phi)e^t\right]},
\end{align}
where, $\delta$ is point on the real coordinate of the complex plane, such that $\delta$ is greater than the real part of all singularities of the integrand.
\begin{eqnarray}
\label{LS}
&L(t,\phi)=12 \phi [(1+1/2 \phi) t+(1+2 \phi)], \nonumber 
\\ 
&S(t,\phi)=(1-\phi)^2t^3+6\phi(1-\phi)t^2+18\phi^2t-
\\
&-12\phi(1+2\phi). \nonumber
\end{eqnarray}
To obtain inverse Laplace transform, Wertheim expanded the denominator of integrand in \eqref{RDF_Wertheim} and applied residue theorem %to series of integrals
\begin{eqnarray}
\label{RDF_Heaviside}
g(r,\phi)=\sum_{n=1}^{\infty}\theta(r-n d)g_n(r,\phi),
\end{eqnarray}
where $\theta(r)$ is the Heaviside step function, and  $g_n(r)$ is result of residue theorem defined at certain shell $n d<r<(n+1) d$. Also, Wertheim obtained $g_1$ in the range $d<r< 2 d$. Then several authors \cite{throop1965numerical, smith1970analytical, henderson1988explicit} extended Wertheim's result to wider range of RDF definitions. 
Later results for RDF in a same form as \eqref{RDF_Heaviside} were obtained using alternative methods by \cite{chang1994real, santos2016exact}.
These results are useful for applications, but are subject to certain limitations. Expressions for RDF defined in a wide range of $r$ are very complicated. As the result, it is not possible to obtain simple analytical expressions for  \eqref{a1}, \eqref{a2} and for the density derivative of RDF. Thus construction RDF in new form allowing, to avoid further numerical calculations is actual problem \cite{henderson2015evaluation, kelly2016analytical}

In case of HS fluid PY theory provides excellent approximation to the exact solution, and it is widely used.  The accuracy problem arise in the cases of small $r$ or large $\phi$. In order to improve PY result, Verlet and Weis \cite{verlet1972equilibrium} proposed an analytical construction, which provides new results within 1\% accuracy \cite{kalikmanov2013statistical}.

In this paper new analytical expression for RDF is obtained. This result makes possible to make explicit integrations \eqref{a1}, \eqref{a2} and derive their analytical expressions. Also analytical expression of the density derivative $\partial g(r,\phi)/\partial \phi$ is obtained. The results of this work are satisfied to the following points:  
\begin{itemize}
	\item HS fluid is considered as reference system for several PT, for this reason, it is necessary to know explicit expression for $g(r)$ over whole range $r>d$.
	%the value and the behavior of $g(r)$ over whole range $r>d$.
	\item The functional form of RDF has to be appropriate for calculation of integrals \eqref{a1}, \eqref{a2} in convenient explicit way.
	\item The minimum of functional for Helmholtz free energy and the pressure
	are obtained by differentiation with respect to $\rho$. Thus, explicit expression for RDF derivatives is needed   
\end{itemize} 

\section{Theory}
\subsection{New form of RDF}
In this section integral \eqref{RDF_Wertheim} is calculated by direct method using residue theorem of complex analysis. For determination of singularity points it is necessary to solve following equation in variable $t$ (denominator of \eqref{RDF_Wertheim} equals to zero):
\begin{eqnarray}
F(t,\phi)=L(t,\phi)+S(t,\phi)e^{t}=0
\end{eqnarray}
after substitution of expressions \eqref{LS}, it transforms into transcendental equation for variable $t$:
\begin{eqnarray}
\label{Eqn1}
12 \phi [(1+1/2 \phi) t+(1+2 \phi)]+[(1-\phi)^2t^3+\nonumber \\
+6\phi(1-\phi)t^2+18\phi^2t-12\phi(1+2\phi)]e^{t}=0.
\end{eqnarray}
Equation \eqref{Eqn1} has infinite number of roots on complex plane. Let us start with obvious root $t=0$ which is pole of the third rang, also it is unique real solution of \eqref{Eqn1}. The other roots are conjugated complex simple poles $t_n=R_n\pm i I_n$, where $R_n$, $I_n$ are real and imaginary parts of complex number. Thus, in accordance to residue theorem,  expression \eqref{RDF_Wertheim} can be rewritten as:
\begin{eqnarray}
	\label{RDF_Residue}
	&g(r,\phi)=1+\dfrac{d}{r}\sum\limits_{\left\lbrace t_n \right\rbrace }\dfrac{t_n L(t_n,\phi)}{12\phi F'(t_n,\phi)}e^{t_n r/d}= \nonumber
	\\
	&=1+\dfrac{d}{r}\sum\limits_{\left\lbrace t_n \right\rbrace}C_n(\phi)e^{t_n r/d},
\end{eqnarray}
where $ F'(t_n,\phi)$ is derivative with respect to $t$ at the point $t=t_n$, here the first term \textquotedblleft1\textquotedblright corresponds to the residue at point $t=0$, the second term is sum over all simple complex poles. Simpler expression can be obtained after summing conjugated poles:
\begin{eqnarray}
\label{RDF_Sum}
&g(r,\phi)=1+\dfrac{2d}{r}\sum\limits_{n=1}^{\infty}A_n(\phi)e^{R_n r/d}\cos(I_n r/d+\alpha_n), \nonumber \\
\end{eqnarray}
where
\begin{eqnarray}
\label{RDF_A_phi}
&A_n=\left|\dfrac{t_n L(t_n,\phi)}{12\phi F'(t_n,\phi)} \right|,\,\, \,\ \alpha_n=\arg\left(\dfrac{t_n L(t_n,\phi)}{12\phi F'(t_n,\phi)}\right). \nonumber 
\\
\end{eqnarray}
Expression \eqref{RDF_Sum} contains only real functions which are depended on $\phi, t_n$. Thus, as one can see from \eqref{RDF_Sum}, in order to calculate RDF, the distributions of roots $t_n(\phi)$ is only needed. 

Let us consider new equation which is the limit $|t|\to \infty$ of equation \eqref{Eqn1}:

\begin{eqnarray}
\label{Eqn2}
12 \phi(1+1/2 \phi)+(1-\phi)^2z^2e^{z}=0
\end{eqnarray}
By introduction of a new variable $q=-\frac{12\phi (1+1/2 \phi)}{(1-\phi)^2}$ it is possible to rewrite \eqref{Eqn2} in more simple form: $z^2e^z=q$. Such equation can be solved exactly in terms of Lambert functions $W(x)$ \cite{scott2006general,corless1996lambertw}
 $$W(x)e^{W(x)}=x.$$

After simple modifications, the above equation can be written as $(2W(x))^2e^{2 W(x)}=4x^2$. Thus, solution of equation \eqref{Eqn2} has the following form 
%\begin{eqnarray}
%\label{Eqn2_System}
%z^2e^z=q \Rightarrow \begin{cases}
%z e^{z y}=\pm q^{1/2} \nonumber \\
%z e^{z (1-y)}=\pm q^{1/2}
%\end{cases}
%\end{eqnarray} 
%where $y$ is a new parameter, which will be determined below. New equations inside the system can be solved by classic Lambert function \cite{corless1996lambertw} $W(x)e^{W(x)}=x$, so \eqref{Eqn2_System} has following view:
%\begin{eqnarray}
%\label{Eqn2_System_Lambert}
%\begin{cases}
%z y = W(\pm q^{1/2} y) \nonumber \\
%z (1-y)= W(\pm q^{1/2} (1-y))
%\end{cases}
%\end{eqnarray} 
%Now it is easy to determinate $y$ from equation 
%$$(1-y)W(\pm q^{1/2} y)=y W(\pm q^{1/2} (1-y)), $$
%since $y=1/2$. And finally solution of \eqref{Eqn2} can be written as:
\begin{eqnarray}
\label{Lambert_0}
z_n=2 W(n,\pm q^{1/2}/2),
\end{eqnarray}
where $n=1,2,...$ enumerates complex branch of Lambert function. 

%Lambert function is known since mid 18-th century from work of Euler. This function is ubiquitous in nature and has applications in many areas of science: for example, in combinatorics for enumeration of trees; calculation of iterated exponentiation; many solutions of dynamical process can be obtained via Lambert function; the similar to the case of this work is solution of linear constant-coefficient delay equations, here the roots of characteristic equation for differential equation are branches of Lambert functions; even in quantum mechanics, an exact solution to the quantum-mechanical double-well Dirac delta function model for equal charges is derived with Lambert function. More detailed information about these applications and properties of Lambert functions can be found in work \cite{corless1996lambertw}. 

\begin{figure}[h]
	\includegraphics[width=8cm]{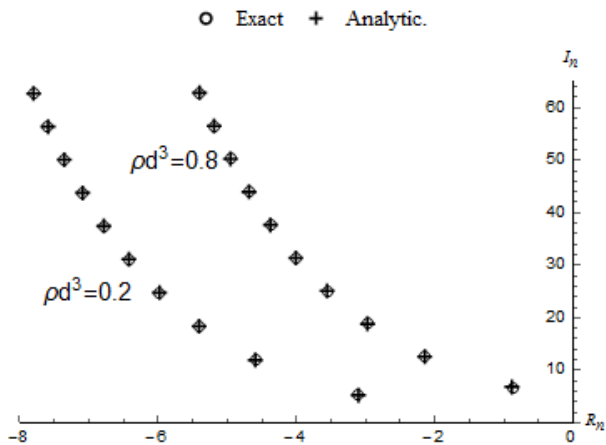}
	\caption{\label{Poles_Distr} Typical distribution of the poles in the upper half complex plane. Circle points correspond to numerical solution \eqref{Eqn1}, cross points are results of analytical expression  with 4 terms \eqref{Lambert_Series}.}
\end{figure}

Using exact solution \eqref{Lambert_0} as the limit, the solution of \eqref{Eqn1} can be written as series of $z_n^{-1}$:
\begin{eqnarray}
\label{Lambert_Series}
t_n=%2 W(n,\pm q^{1/2}/2) 
z_n+\sum_{k=1}^{\infty}a_k z_n^{-k}
\end{eqnarray}
where coefficients $a_n$ depend only on density $\phi$ and can be found after substitution of \eqref{Lambert_Series} in \eqref{Eqn1}. In this work four terms were calculated, in Fig.~\ref{Poles_Distr} one can see typical distribution of the poles on upper half complex plane (down half plane looks same, but symmetry reflected). From this figure and Table~\ref{Poles_Table}, one can see that analytical expression with 4 terms \eqref{Lambert_Series} is  accurate approximation for numerical solutions of \eqref{Eqn1}.

\begin{table}
	\caption{\label{Poles_Table} Comparison of real $R_1$ and imaginary $I_1$ parts of first root $t_1$ which are calculated by numerical results for \eqref{Eqn1} (\textquotedblleft Exact\textquotedblright-column) and by analytical expression \eqref{Lambert_Series} with 4 terms (\textquotedblleft Analytic.\textquotedblright-column). Results are obtained for several densities, 	relative deviation is not biger than 1\%.}
	\begin{ruledtabular}
	\begin{tabular}{cccccccc}
	$\rho d^3$	& Exact $R_1$ & Exact $I_1$ & Analytic. $R_1$ &  Analytic. $I_1$ \\
	\hline
	0.1	& -4.072 & 4.761 &	-4.072 & 4.761 \\
	0.3 & -2.487 & 5.398 & -2.482 & 5.396 \\
	0.5	& -1.667 & 5.889 & -1.650 & 5.901 \\
	0.7 & -1.089 & 6.351 & -1.079 & 6.418

	\end{tabular}
\end{ruledtabular}
\end{table}

For the aims of this work it will be enough to consider only one term in the sum \eqref{Lambert_Series}. Here and below following expression is used as approximations for the roots of \eqref{Eqn1}:
\begin{eqnarray}
\label{Lambert}
t_n\simeq z_n+\frac{2(1-5\phi+5\phi^2)}{(1-\phi)(2+\phi)}z_n^{-1}
\end{eqnarray}  

It is easy to verify, that for all poles $R_n=Re(t_n)<0$, and $R_{n+1}<R_n$, then contribution of exponential n-th term in sum \eqref{RDF_Sum} rapidly decreases, when the number $n$ increases. Thus, for accurate result summation over all terms in \eqref{RDF_Sum} is not required and \eqref{RDF_Sum} can be rewritten as analytical expression with $M$ terms: 
\begin{eqnarray}
\label{RDF_PY_L}
g(r)=1+\frac{2d}{r}\sum_{n=1}^{M}A_n(\phi)e^{R_n r/d}\cos(I_n r/d+\alpha_n).
\end{eqnarray}

\begin{figure}[h]
	\includegraphics[width=8cm]{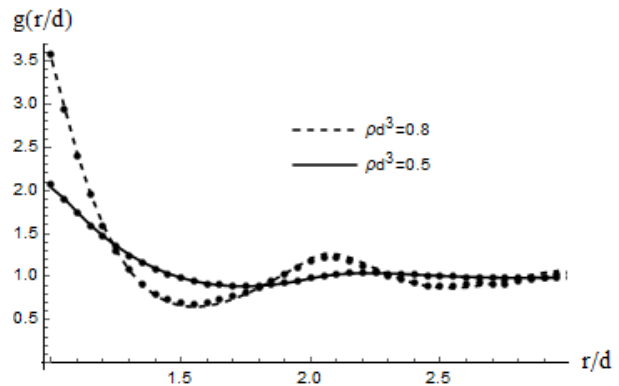}
	\caption{\label{RDF_PY} PY RDF as function of dimensionless distance $r/d$ calculated for two cases of density $\rho d^3=0.8, \rho d^3=0.6$, dashed and solid lines, respectively. Points correspond to published numerical data \cite{throop1965numerical}.}
\end{figure}

Accurate result for the point of contact ($r=d$) requires summation over large number of terms (poles). This effect is  well known in inverse Laplace or Fourier transforms and is called  \textquotedblleft Gibbs phenomenon\textquotedblright. Non-formally there is the overestimate in value of finite series near the point of function jump ($g(r)$ near $r=d$). This artifact appears when discontinuous function approximated by finite series of continuous functions. Amplitude of the overestimate does not disappear as $M$ increases, and tends to finite value. However infinite limit of series does not the overestimates, because the location of the overestimate moves aside point of function discontinuity. In other words, there is pointwise convergence, but not uniform convergence \cite{pinsky2002introduction}. Practical result can be found as $g(d)\simeq g(d+\epsilon(M))$, where $\epsilon\ll d$ and $\epsilon\to0$ when $M\to\infty$. Solid and dashed curves in Fig.~\ref{RDF_PY} correspond to $M=1000$. In this figure one can see, that comparison of analytical expression \eqref{RDF_PY_L} and numerical results \cite{throop1965numerical} demonstrates good accuracy. For further calculations it is enough to use only $M=100$ terms in \eqref{RDF_PY_L}.

In spite of wide applications, PY approximation has two weak points: the contact value $g(d)$ is too low at high density; phase of oscillation at large distance $r$ differs from the exact one. In order to correct these artifacts construction of Verlet-Weis can be used \cite{verlet1972equilibrium}. This is achieved by introduction of a modified density $\phi_m=\phi+\phi^2/16$ and modified HS diameter $d_m=\left(\phi_m/\phi\right)^{1/3}d$ in order to correct RDF oscillation. Verlet and Weis, also, proposed an addition term which improved contact value $g(d)$, so corrected RDF has following form:
\begin{eqnarray}
\label{RDF_VW}
&g^{VW}(r/d)=g(r/d_m;\phi_m)+\dfrac{1}{r}e^{\frac{\alpha(r-d)}{d}}\cos\dfrac{\alpha (r-d)}{d}, \nonumber \\
\end{eqnarray}
where parameters $A$ and $\alpha$ can be found from
\begin{eqnarray}
&\dfrac{A}{d}=\dfrac{3}{4}\dfrac{\phi_m(1-0.7117 \phi_m-0.114 \phi_m^2)}{(1-\phi_m)^4} \nonumber \\
&\alpha = \dfrac{24 A/d}{\phi_m g(d_m, \phi_m)} \nonumber
\end{eqnarray}
The form of expression of added term in \eqref{RDF_VW} coincides with analytical result \eqref{RDF_Sum}. This fact helps the process of further calculations of corrected form \eqref{RDF_VW}.

\begin{figure}[h]
	\includegraphics[width=8cm]{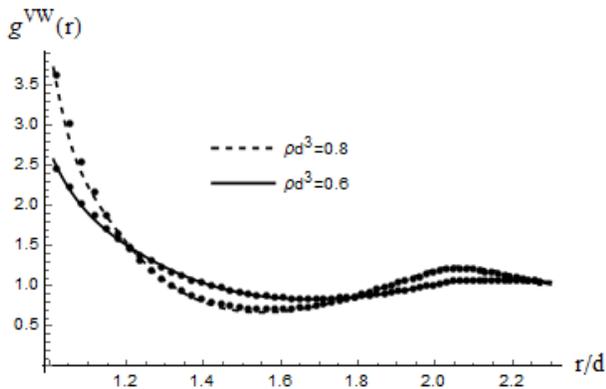}
	\caption{\label{Fig_RDF_VW} Corrected RDF \eqref{RDF_VW} as function of dimensionless distance $r/d$ calculated for two cases of density: $\rho d^3=0.8, \rho d^3=0.6$, dashed and solid lines, respectively. Points correspond to published data of Monte Carlo simulations \cite{barker1971monte}}
\end{figure}

After application of VW procedure the obtained RDF can be compared with the results of HS modeling. Fig.~\ref{Fig_RDF_VW} demonstrates, that analytical expression  \eqref{RDF_VW} for M=100 with VW corrections correctly describes the behavior of HS fluid.

It is important to note that approximated expressions for singularities of \eqref{RDF_Wertheim} were investigated in \cite{perram1980role}. However, in previous works exact solution of \eqref{Eqn2} has never been implemented. The exact solution of \eqref{Eqn2} is required to obtain right expression for derivative $\partial g(r,\phi)/\partial \phi$.

\subsection{Density derivative of RDF}

The method proposed in this work allows to obtain the density derivative of RDF without decrease of accuracy. It follows from the fact that all variables of \eqref{RDF_Sum} are explicit functions of density, so all partial derivatives can be written as analytical expressions.
\begin{eqnarray}
\frac{\partial g(r,\phi)}{\partial \phi}=\frac{1}{r}\sum_{n}\left[\frac{\partial C_n}{\partial \phi}+\frac{\partial C_n}{\partial t_n}\frac{\partial t_n}{\partial \phi}+r C_n \frac{\partial t_n}{\partial \phi} \right] e^{t_n r}, \nonumber
\end{eqnarray}
where derivative $\partial t_n/\partial \phi $ can be calculated exactly using properties of Lambert function \cite{corless1996lambertw} 
$$
\frac{\partial W_n(x)}{\partial x}=\frac{1}{x}\frac{W_n(x)}{W_n(x)+1}.
$$
This relationship is correct for any branch of Lambert function. Thus,
\begin{eqnarray}
&\dfrac{\partial z_n}{\partial \phi}=\dfrac{1}{q}\dfrac{W_n(\pm q^{1/2}/2)}{W_n(\pm q^{1/2}/2)+1}\frac{\partial q}{\partial \phi}= \nonumber \\ &=\dfrac{-1-2\phi}{\sqrt{6}(-\phi(2+\phi))^{3/2}}\dfrac{z_n}{z_n+2}. \nonumber
\end{eqnarray}
 
\subsection{Perturbation terms}

Analytical expressions of RDF play important role in calculation of perturbation terms. Indeed, numerical integrations of \eqref{a1} and \eqref{a2} at each density is inconvenient. The derived above expression for RDF of HS makes it possible to obtain the perturbation terms in analytical form. Also analytical integrated expressions \eqref{a1}, \eqref{a2} can be used in equation of state for real fluid and density functional theory calculations.

\begin{figure}[h]
	\includegraphics[width=8cm]{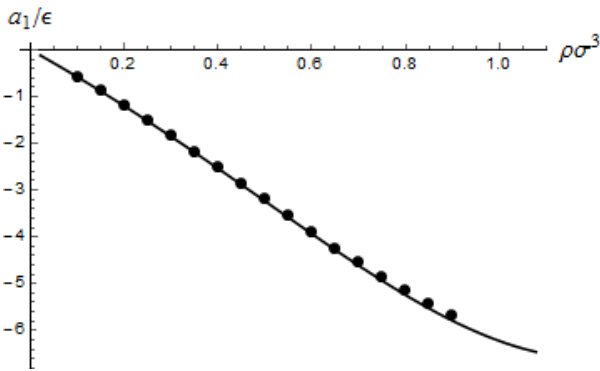}
	\caption{\label{Fig_a1} The density $\rho \sigma^3$ dependence of the first perturbation terms for the LJ system at temperature $T=\epsilon/k$. 
	Solid curve is analytical result \eqref{a1_result}, dots correspond to Monte Carlo simulations \cite{lafitte2013accurate}}.	
\end{figure}

Let us consider the system of molecules interacting by potential
$$
U(r)=\epsilon \gamma \left[ \left(\frac{\sigma}{r}\right)^{\lambda_r}-\left(\frac{\sigma}{r}\right)^{\lambda_a}\right],
$$
where $\epsilon$ is characteristic energy, $\gamma=\frac{ \lambda_r}{\lambda_r-\lambda_a}\left(\frac{\lambda_r}{\lambda_a}\right)^{\frac{\lambda_r}{\lambda_r-\lambda_a}}$ is a constant, which in case of Lennard-Jones (LJ) fluid ($\lambda_r=12, \lambda_a=6$) equals to $\gamma=4$. Then, at certain temperature $T$ for Barker-Henderson PT the reference system is a system of HS molecules with diameter $d(T)<\sigma$. Thus, the first two terms of PT are defined by RDF of HS molecules with known diameter $d$. Using the explicit spatial dependence of RDF $g^{VW}(r)$ \eqref{RDF_VW}, all necessary integrals can be expressed in the following general form:
\begin{eqnarray}
\int_{x_0}^{\infty}\frac{dr}{r^n}e^{a r}=x_0^{-n+1}E_n(- a x_0)
%(-a)^{1-n}\Gamma[1-n, - a x_0]
\end{eqnarray}  
where $E_n(x)$ is the exponential integral \cite{gradshteyn2014table}.
After substitution of RDF $\eqref{RDF_VW}$ into \eqref{a1} and \eqref{a2}, the first and the second terms of PT are
\begin{widetext}
	\begin{eqnarray}
	\label{a1_result}
	&a_1(\phi)= 12 \gamma \epsilon \phi \left\lbrace 
	\dfrac{x_0^3}{\lambda_r-1}+\dfrac{x_0^3}{\lambda_a-1}
	+\dfrac{x_0^2}{k}
	\sum_{n} C_n \left[E_{\lambda_r-1}(-k x_0 t_n)
	+E_{\lambda_a-1}(-k x_0 t_n)\right]+
	\right.
	\\
	& \left. 
	+ \dfrac{x_0^2 A}{2} \left[E_{\lambda_r-1}(-x_0\tau) +E_{\lambda_r-1}(-x_0\tau^*) +E_{\lambda_a-1}(-x_0\tau)+E_{\lambda_a-1}(-x_0\tau^*)\right] 
	\right\rbrace \nonumber
	\\\nonumber
	\end{eqnarray}
\begin{eqnarray}
	\label{a2_result}
	&a_2(\phi)= -6 \gamma^2 \epsilon^2\phi K %(1+\chi)
	\bigg\{%\left\lbrace 
	\dfrac{x_0^3}{2\lambda_r-1}-\dfrac{2 x_0^3}{\lambda_r+\lambda_a-1}+\dfrac{x_0^3}{2\lambda_a-1}+\nonumber
	\\ 
	&+\dfrac{x_0^2}{k}
	\sum_{n}C_n \left[E_{2\lambda_r-1}(-k x_0 t_n)
	-2E_{\lambda_r+\lambda_a-1}(-k x_0 t_n)+E_{2\lambda_a-1}(-k x_0 t_n)\right]+%\right.
	\\
	& + \dfrac{x_0^2 A}{2} \left[E_{2\lambda_r-1}(-x_0\tau) +E_{2\lambda_r-1}(-x_0\tau^*)-2E_{\lambda_r+\lambda_a-1}(-x_0\tau)- 
	\nonumber \right.
	\\
	&\left.-2E_{\lambda_r+\lambda_a-1}(-x_0\tau^*) +E_{2\lambda_a-1}(-x_0\tau)+E_{2\lambda_a-1}(-x_0\tau^*)\right]\bigg\}\nonumber
	\end{eqnarray}
\end{widetext}
where $\tau=-\alpha+i\alpha$ and conjugated one $\tau^{*}=-\alpha-i\alpha$. Taking into account expression \eqref{Lambert}, results \eqref{a1_result},\eqref{a2_result} are analytical expressions with explicit dependence on the density $\phi$. Fig.~\ref{Fig_a1} shows the dependence on density $\rho \sigma^3$ of $a_1$ for LJ fluid at temperature  $T=\epsilon/k$ . As one can see, analytical expression \eqref{a1_result} (solid line) demonstrates excellent agreement with Monte Carlo simulations (dots) \cite{lafitte2013accurate}.

\begin{figure}[h]
	\includegraphics[width=8cm]{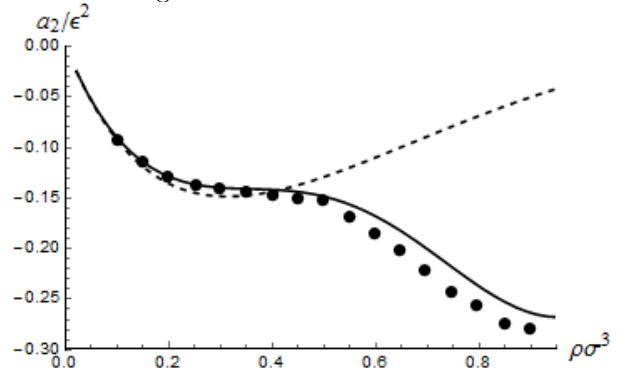}
	\caption{\label{Fig_a2} The density $\rho \sigma^3$ dependence of the second perturbation terms for the LJ system at temperature $T=\epsilon/k$. Dashed curve corresponds to analytical result \eqref{a2_result} without correction prefactor $\chi=0$. Solid curve is analytical result \eqref{a2_result} taking into account correction prefactor from work \cite{lafitte2013accurate}. Dots correspond to Monte Carlo simulations \cite{lafitte2013accurate}.}
\end{figure}
In case of the second term $a_2$, the comparison with simulation data is more complicated. Definition \eqref{a2} is approximated result and corresponds to \textquotedblleft compressibility approximation\textquotedblright of BH. How one can see from Fig.~\ref{Fig_a2}, result \eqref{a2_result} without correction prefactor (dashed line) works well only for low range of densities. Better version of correction prefactor $1+\chi$ was implemented in work \cite{lafitte2013accurate}. With account of this $\chi$ from work \cite{lafitte2013accurate}, expression \eqref{a2_result} matches numerical data in whole range of the densities, solid line in Fig.~\ref{Fig_a2}.

\section{Conclusion}
In this work analytical expression for inverse Laplace transform of Wertheim's result was calculated. This expression was represented as the sum of the residues at the simple poles. Exact expressions for the complex poles were found in terms of Lambert function. Accurate results were obtained by appropriate approximations for poles and the overall sum. After Verlet-Weis corrections this result can be applied to description of hard sphere fluid. Also the density derivative, and two first terms of Barker-Henderson perturbation theory were calculated. All obtained results coincide well with published results of Monte Carlo simulations. Results of this work have important implications for modeling of real fluids by density functional methods and perturbation theories.
\bibliography{newHSRDF}
\end{document}